\documentclass[%
 reprint,
 amsmath,amssymb,
 aps,
pre,
]{revtex4-1}
\usepackage{graphicx}
\usepackage{dcolumn}
\usepackage{bm}
\usepackage{float}
\usepackage{caption}
\usepackage{subfig}
\usepackage{hyperref}
\hypersetup{colorlinks,breaklinks,linkcolor=blue,urlcolor=blue,
anchorcolor=blue,citecolor=blue}
\usepackage{color}
\usepackage{natbib}

\begin{document}


\title{Nonlinear dynamics and band transport in a superlattice driven by a plane wave}

\author{A.~Apostolakis$^1$, M. K.~Awodele$^1$, K. N.~Alekseev$^1$, F. V.~Kusmartsev$^1$, A. G.~Balanov$^{1,2}$}
\affiliation{
$^1$Department of Physics, Loughborough University, Loughborough LE11 3TU, United Kingdom\\
$^2$Saratov State Technical University, Politechnicheskaja 77, Saratov, 410054, Russia
}

\begin{abstract}
A quantum particle transport induced in a spatially-periodic potential by a propagating plane wave  
has a number important implications in a range of topical physical systems. Examples include acoustically driven semiconductor superlattices and cold atoms in optical crystal.  
Here we apply kinetic description of the directed transport in a superlattice beyond standard linear approximation,  and utilize exact path-integral solutions of the semiclassical transport equation.
We show that the particle drift  and average velocities have non-monotonic dependence on the wave amplitude with several prominent extrema. Such nontrivial kinetic behaviour is related to global bifurcations developing with an increase of the wave amplitude. They cause dramatic transformations of the system phase space and lead to changes  of the transport regime.
We describe different types of phase trajectories contributing to the directed transport and analyse their spectral content.
\end{abstract}


\date{\today}

\maketitle

\section{\label{sec:level1}INTRODUCTION}

Semiclassical  models are widely used to describe particle transport in quantum  systems with spatially periodic potential, such as
high-frequency semiconductor devices 
\cite{esaki1970superlattice,BASS1994219,ignatov1995thz}, cold atoms in optical potential 
\cite{greenaway2013resonant}, photonic crystals \cite{Wilk2003} and waveguide arrays \cite{Longhi2007}. 
Being realized in the systems with inherent semiclassical
nonlinearities this transport is often accompanied by a development of instabilities 
and catastrophes \cite{Kolovsky09,kusmartsev1989application}, which can be utilized in applications.
In particular, electron transport in semiconductor superlattices (SLs) is associated with a variety of 
quantum mechanical and kinetic effects \cite{BASS1986237, Wacker20021}.
Those effects are able to enhance the electron mobility and induce  terahertz (THz) dynamics of electrons \cite{esaki1970superlattice,leo1992observation,ignatov1991transient} making  SLs  to be a promising element for designing THz 
sources \cite{Waschke93,SankinAPL16}, amplifiers \cite{hyart2009terahertz}, and frequency mixers \cite{Pavelyev14}.

During the last decades a number of interesting and important phenomena associated with interaction of carriers in SLs with high-frequency coherent
phonons have been discovered \cite{Bartels98,Cav02,beardsley2010coherent,maryam2013dynamics,Shinokita16}. The characteristic feature of such systems is that in the case  of the acoustic wave the propagation effects cannot be neglected, as it typically done when electromagnetic waves are considered. 
In particular, it was found out that the SLs are able to amplify the hypersonic acoustic waves  by mechanisms involving either the interwell tunnelling \cite{beardsley2010coherent} or the stimulated Cherenkov effect \cite{Shinokita16}.
It was also recently shown that acoustic stimuli can be utilised as a powerful mean to induce and to control high-frequency transport and  resulting emission of electromagnetic waves in semiconductor heterostructres \cite{fowler2008semiconductor,Bru2012,POY2015}. 
Although a certain progress has been achieved in understanding of the related high-frequency electro-acoustic phenomena
\cite{Shmelev88,greenaway2010using,Bau2013121}, 
the underlying physical mechanisms and the related nonlinear dynamics are still poorly studied \footnote{Earlier nonlinear electron dynamics in a miniband SL driven by a plane wave has been studied in \cite{BassPRB95}. That work, however, is focused on the specific case of a very high-frequency wave.}.

In this paper we theoretically study semiclassical dynamics of particles in a spatially periodic potential induced by a propagating plane wave on a practically motivated example of  a single miniband SL driven by a high-frequency acoustic plane wave 
\cite{Shinokita16}.
In order to describe the interaction between electrons and acoustic wave in crystals one has conventionally used the linearized transport equations  \cite{Spector62,Kazarinov63}. On the contrary, here we use the exact path integral solution of the Boltzmann transport equation for an arbitrary wave amplitude \cite{Chambers52,Budd63-PI,maccallum1963kinetic}. 
Remarkably, this nonperturbative approach requires knowledge about Hamiltonian dynamics of electrons, which in our case of plane wave is strongly nonlinear.
We characterise the directed transport in the SL by introducing time-averaged velocity of charge carriers [Eq.~\ref{eq:v_m}]
and their drift velocity.
The latter takes into account the scattering events.
Both characteristics demonstrate nontrivial dependencies upon the amplitude of the acoustic wave [Fig.~\ref{fig1}]. 
 
Depending on the wave parameters, different dynamical regimes are realised in the system, each represented in the phase space by a characteristic phase portrait [Fig.~\ref{fig:ph_sp}]. 
We reveal and classify various trajectories, which affect charge transport, and establish a series of global bifurcations associated with dramatical restructuring of the phase space [Eq.~(\ref{eq:u_cr})]. Here the wave amplitude serves as a control parameter determining sharp transitions between different dynamical regimes at the bifurcations. Importantly, we find out that these transitions evoke the characteristic changes in the averaged velocities of the carriers [Fig.~\ref{fig1}].  

Before the first bifurcation the dominant dynamical regime is a nonlinear
dragging of particles [Eq.~(\ref{eq:pend})], which is characteristic for an ordinary acousto-electric effect \cite{Parmenter53}.
Beyond the first and successive bifurcations we observe a switching to the dynamical regimes termed nonlinear Bloch-like oscillations 
\cite{greenaway2010using}. These complex Bloch oscillations are characterized by specific quasi-periodic motion in the phase plane [Fig.~\ref{real_tr}(c)]. The related trajectories drift in the direction opposite to the propagating wave, which can lead to the appearance of negative time-averaged electron velocity [Fig.~\ref{fig1}(b)].
We also investigate how the bifurcations change the spectral content of the averaged electron trajectories, which is used for understanding an unique high-frequency response of SLs to hypersonic excitations [Eq.~(\ref{eq:cutoff})]. 

The paper has the following structure. Section \ref{sec:level2} 
is devoted to a semiclassical formalism for characterisation of charge transport in SL at presence of scattering.
In Section \ref{sec:level3} different dynamical regimes associated with charge transport are revealed, and bifurcation transitions between them are studied. Section \ref{sec:level4} is dedicated to the spectral analysis of different trajectories. Finally, in Section \ref{sec:level5} we summarize results and 
discuss various applications.

\section{\label{sec:level2}SEMICLASSICAL DYNAMICS AND DIRECTED TRANSPORT}

\begin{figure}
\includegraphics[scale=0.7]{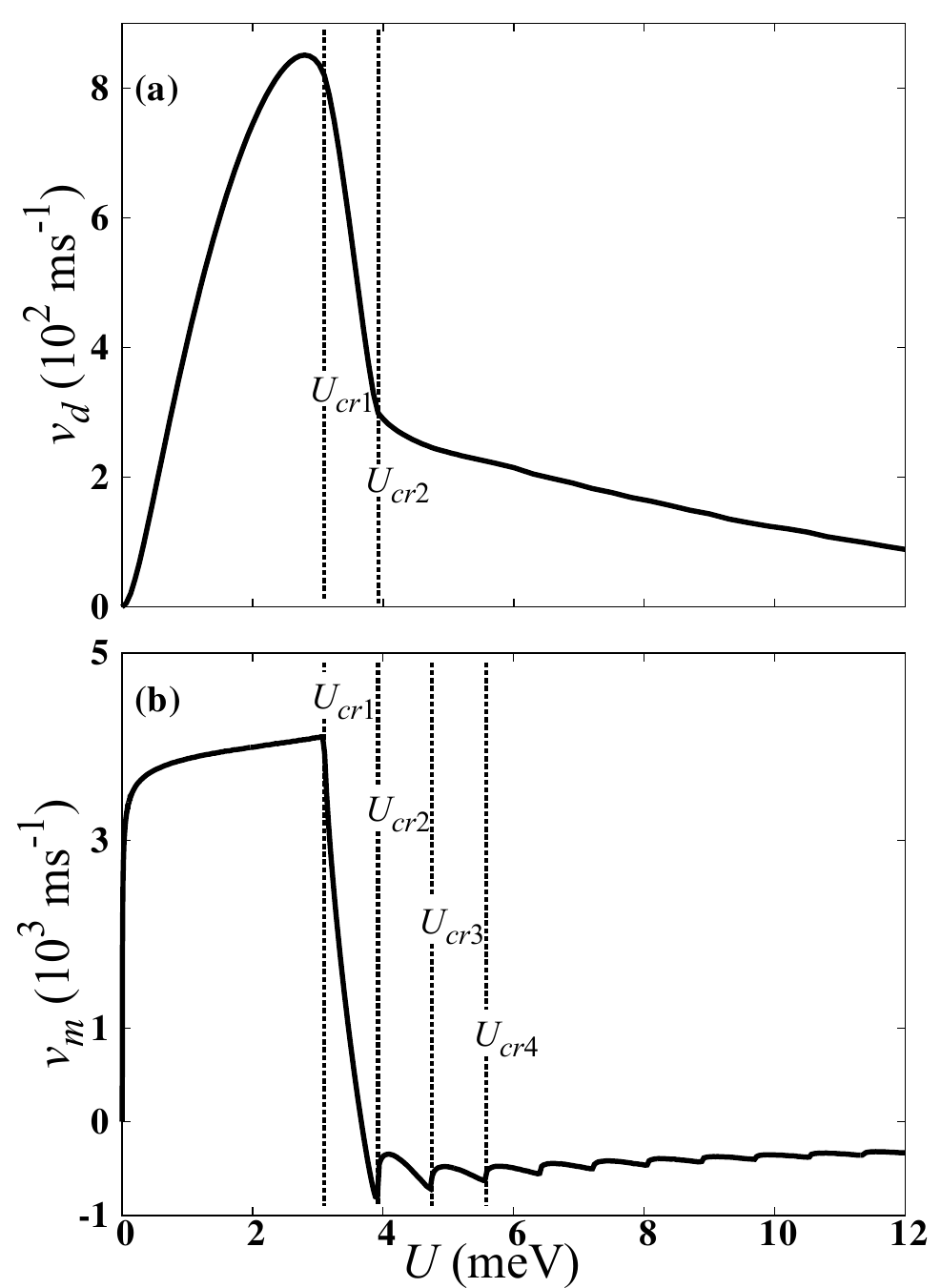}
   \caption{(a) The drift velocity  $v_d$ as a function of the acoustic wave amplitude $U$; (b) the dependence of the time--averaged electron velocity $v_m$ upon $U$. The vertical lines correspond to the critical values of $U$.} 
\label{fig1}
\end{figure}

We consider the semiclassical dynamics of an electron, which tunnels through spatio-temporal potential of strongly-coupled SL driven by a plane acoustic wave.  The effect of a plane wave can be represented  by a moving potential $V(x,t)=-U\sin[(k_s(x+x_0) +\omega_s t)]$ \cite{Kazarinov63,greenaway2010using}, which, as time $t$ changes, propagates along SL's axis in $x$-direction. The
wave amplitude $U$ depends on strain and the deformation potential \cite{greenaway2010using}, $k_s=\omega_s/v_s$ is the wave number, $\omega_s$ is the wave frequency, and $v_s$ is the speed of sound in the materials of SL. Displacement $x_0$ defines the initial phase of the driving wave.

The election transport is assumed only within the lowest miniband, and interminiband tunnelling is neglected. Then, within the tight-binding approximation, the kinetic energy of electron is defined as $E(p_x) = \Delta/2\left[1-\cos(p_xd/\hbar)\right]$ \cite{esaki1970superlattice}, where $p_x$ is the electron quasi-momentum, $\Delta$ is the miniband width, and  $d$ is the period of SL.
The semiclassical Hamiltonian $H(x, p_x)=E(p_x)+V(x,t)$ yields the following equations of  motion:
\begin{subequations}
\label{eq:all-dot}
\begin{eqnarray}
v_x=\frac{dx}{dt}&=&\frac{\partial H}{\partial p_x }=\frac{\Delta d}{2\hbar}\sin\frac{p_xd}{\hbar},  \label{xdot}\\
\frac{dp_x}{dt}&=&-\frac{\partial H}{\partial x } =k_sU\cos[(k_s(x+x_0) -\omega_s t)] \label{pdot}.
\end{eqnarray}
\end{subequations}
In order to characterise the particle transport described by model (\ref{eq:all-dot})
in the presence of scattering we introduce the drift velocity of electrons $v_d$.  Earlier in \cite{greenaway2010using} $v_d$ was calculated using 
the Esaki-Tsu formalism \cite{esaki1970superlattice}.
In this case the drift velocity depends on the initial conditions, and will be different for different trajectories. Therefore, this approach does not allow to examine generic transport characteristics, and to reveal global instabilities (bifurcations), which can be developed in the system. 

Here we employ a more generic approach by using the time-dependent path integral as a steady solution of the time-dependent Boltzmann transport equation 
\cite{Budd63-PI,maccallum1963kinetic,bass1981theory}.  
Within this framework the drift velocity is defined as
\begin{equation} 
 v_d=\int_0^T\frac{dt}{T}\int_{-\infty}^t e^{\frac{-(t-t_0)}{\tau}}v_x(t,t_0)\frac{dt_0}{\tau} \label{Boltz_vd}, 
\end{equation}
where $t_0$ is the moment of time, when the electron can be found at position $x_0$; 
$\tau$ is the scattering time of electrons, and $T=2{\pi}/{\omega_s}$ is the period of the acoustic plane wave.
 
This implies that in order to find $v_d$, we need to know the nonlinear dynamics of electrons governed by Eqs.~(\ref{eq:all-dot}). 
Note that  (\ref{Boltz_vd}) can be considered as an extension of  the seminal
Chambers result \cite{Chambers52}, and was earlier used for consideration of the charge transport in SL with ac electric filed applied \cite{IgnatovRomanov76,winnerl1997quasistatic}, also in the presence of a static magnetic field \cite{Hyart09-magfld}.
According to Eq.~(\ref{Boltz_vd}) the drift velocity can be understood as a velocity of an electron averaged over all initial moments $t_0$ and over the time period $T$  after taking into account a probability of electron scattering within the time interval between $t-t_0$ and $t-t_0+dt$. 

There is an alternative way to introduce the drift velocity by averaging the electron velocities not over initial time $t_0$, but across the initial positions $x_0$ or, equivalently, across the initial phases of the acoustic wave. Since the Hamiltonian $H$ is periodic in time, the velocity of electrons at the presence of scatterings $v(x,t)$ can be expanded in Fourier series 
\cite{bass1981theory} as
\begin{equation}
v(x,t)=\sum_n v_n e^{ink_s(x-v_st)} \nonumber
\end{equation}
with
\begin{eqnarray}
v_n=\frac{1}{\lambda}\int_0^{\lambda}d{x_0} \int_0^{\infty} e^{\frac{-t{'}}{\tau}}v_x(x_0,t') \nonumber \\
\times e^{-ink_s(x_0+v_st')}\frac{dt^{'}}{\tau}, \label{vd_vn}
\end{eqnarray}
where $\lambda=2\pi/k_s$ represents the wavelength of the propagating wave. By applying Jacobian $J=\partial(tv_s,t-t_0)/\partial(t,t_0)$, the set of variables $(x_0,t')$  can be substituted by $(t_0, t)$, for which Eq. (\ref {vd_vn}) takes the form
\begin{equation}
v_n=\frac{1}{T}\int_0^{T}d{t} \int_{-\infty}^t e^{\frac{-(t-t_0)}{\tau}}v_x(t,t_0)e^{-in\omega_st}\frac{dt_0}{\tau}.
\label{eq:vn_t}
\end{equation}
Remarkably, the zeroth Fourier component (for $n=0$) of the velocity in  (\ref{eq:vn_t}) is identical to the drift velocity $v_d$ calculated with the time dependent path integral (\ref{Boltz_vd}), thus implying that integration over all electron initial positions $x_0$ realized in  (\ref{vd_vn})  is an equivalent to integration over all  
starting times $t_0$  in (\ref{Boltz_vd}). Later we will use this equivalence to better understand the contribution from different types of the electron trajectories to the transport characteristics  of the system.   In addition, this is  used for an optimization of numerical calculation of the drift velocity, since the direct averaging over the initial positions is easier to implement.
In our study we chose the parameters of a realistic SL \cite{fowler2008semiconductor,greenaway2010using}, namely: $\Delta=7$ meV, $d=12.5$ nm, $v_s=5000$ m/s, $\tau=250$ fs and $\omega_s = 4\times10^{11}$ rad/s. However, we note that the phenomena discussed can be found in a wide range of the parameter values.

Figure \ref{fig1}(a) illustrates the change of drift velocity $v_d$  with variation of the acoustic wave amplitude $U$. The drift velocity was estimated for the electrons starting at the time moment $t_0$ with $p_x=0$, which corresponds to the experimental conditions of the non-degenerate electron gas close to zero temperature. One can see that the dependence $v_d(U)$ is strictly non-monotonic, and can be characterized by two representative values of $U$, which correspond to the maximum of $v_d$ (value $U_{cr1}$) and to the point, where the curve sharply changes its slop (value $U_{cr2}$). 
The presence of a prominent maximum reminds the classical Esaki-Tsu  $v_d(F)$ dependence \cite{esaki1970superlattice}, which reflects  the effect of a constant electric field $F$ on the drift velocity $v_d$. In this case, the dominant transport regime 
is related to the conventional Bloch oscillations 
\cite{Lyssenko-PRL97}, and existence of the maximum in the  $v_d(F)$ dependence is evoked by multiple scattering events. 
However, as it was found in 
\cite{greenaway2010using}, in the case of the acoustically driven SL the trajectories of the electrons  and, correspondingly, the transport regime can dramatically change with variation of $U$.
In order to understand what causes the non-monotonic character of the dependence shown in Fig. \ref{fig1}(a), the scattering events or the changes in dynamics, we calculated the mean velocity of electrons $v_m$ averaged over the time: 
\begin{equation} 
v_m=\frac{1}{\lambda}\int_0^{\lambda}d{x_0} \int_0^{\Delta t} v_x(t+t_0,t_0)\frac{dt}{\Delta t}.
\label{eq:v_m}
\end{equation}
In fact, formula (\ref{eq:v_m}) is a finite time version of the Eq. (\ref{vd_vn}) under assumption that $\tau\to\infty$. In our calculations we chosen $\Delta t$=2 ns, which is large enough to ensure convergence in the numerical calculations of (\ref{eq:v_m}).  Note that $v_m$ is an important transport characteristic for cold atom systems, whose dynamics can be also described using spatially-periodic Hamiltonians  \cite{greenaway2013resonant}. The dependence of $v_m(U)$ is shown in Fig. \ref{fig1}(b). Comparison of Fig. \ref{fig1}(a) and (b) reveals that both graphs demonstrate their characteristic features at almost the same values of $U$ (indicated by dashed lines). Namely, they both demonstrate prominent maxima at the values $U=U_{cr1}$, and have prominent features at $U=U_{cr2}$. In addition, the dependence $v_m(U)$ has multiple maxima, and in contrast to $v_d$,  $v_m$ can attain negative values. All this evidences that specific changes in $v_d$ with variation of $U$ are  associated with the transitions between different  dynamical regimes in electron transport in the SL.

\section{\label{sec:level3} TRAJECTORIES AND BIFURCATIONS}

\subsection{\label{subsec:level-A} Dynamical regimes, phase portraits and bifurcations}

To understand how dynamics of electrons affects $v_m$ and $v_d$ we analyse the equations of motion 
in the moving reference frame $x'(t)=x(t)+x_0-v_st$.
In this case the model has the form 
\begin{subequations}
\label{eq:all-prime}
\begin{eqnarray}
\dot{x}'&=&v_0\sin\frac{p_xd}{\hbar} -v_s, \label{eq:xprime}\\ 
\dot{p}_x&=&k_sU\cos(k_sx'), \label{eq:pprime}
\end{eqnarray}
\end{subequations}
which, in contrast to Eqs.~(\ref{eq:all-dot}) 
does not explicitly depend on the time.
Here, $v_0=\Delta d/(2\hbar)$ characterises the maximal possible change of electron velocity $v(p_x)$ within the miniband.  
For our choice of SL parameters  $v_0=6.6\times 10^4$ m/s and the key ratio $v_0/v_s$ is $\approx 13$.
New equations of motion correspond to the Hamiltonian $H'=E'(p_x)+V(x')$
with the modified energy dispersion relation and time-independent potential energy
\begin{equation}
E'(p_x)=E(p_x)-v_s p_x, \quad V(x')=-U\sin(k_sx'). \label{eq:Eprime}
\end{equation}
%
It is worth noting that in the limit $v_s\rightarrow 0$ the Hamiltonian $H'$ transforms to well-known Hamiltonian of the  
classical 2D Harper model \cite{Harper55,Geisel93}.

For both numerical and analytical studies it is convenient to rewrite Eqs.~(\ref{eq:all-prime}) 
in a dimensionless form
\begin{subequations}
\label{eq:all-tilde}
\begin{eqnarray}
\frac{d\tilde{x}}{d\tilde{t}}&=&\frac{v_0}{v_s}\sin\tilde{p} -1,  \label{eq:xtilde}\\
\frac{d\tilde{p}}{d\tilde{t}}&=&\frac{Ud}{\hbar v_s}\cos\tilde{x},  \label{eq:ptilde}
\end{eqnarray}
\end{subequations}
where $\tilde{x}=k_sx'$, $\tilde{p}=p_xd/\hbar$, and $\tilde{t}=\omega_st$.  These normalized quantities have clear physical interpretations. Namely, $\tilde{p}$ corresponds to the phase of the de Broglie wave characterising a position of the electron within 
Brillouin zones, 
whereas  $\tilde{x}$ and $\tilde{t}$ are associated with 
the phase and time components of the total phase of the acoustic wave, respectively. 

\begin{figure}
  \includegraphics[scale=0.85]{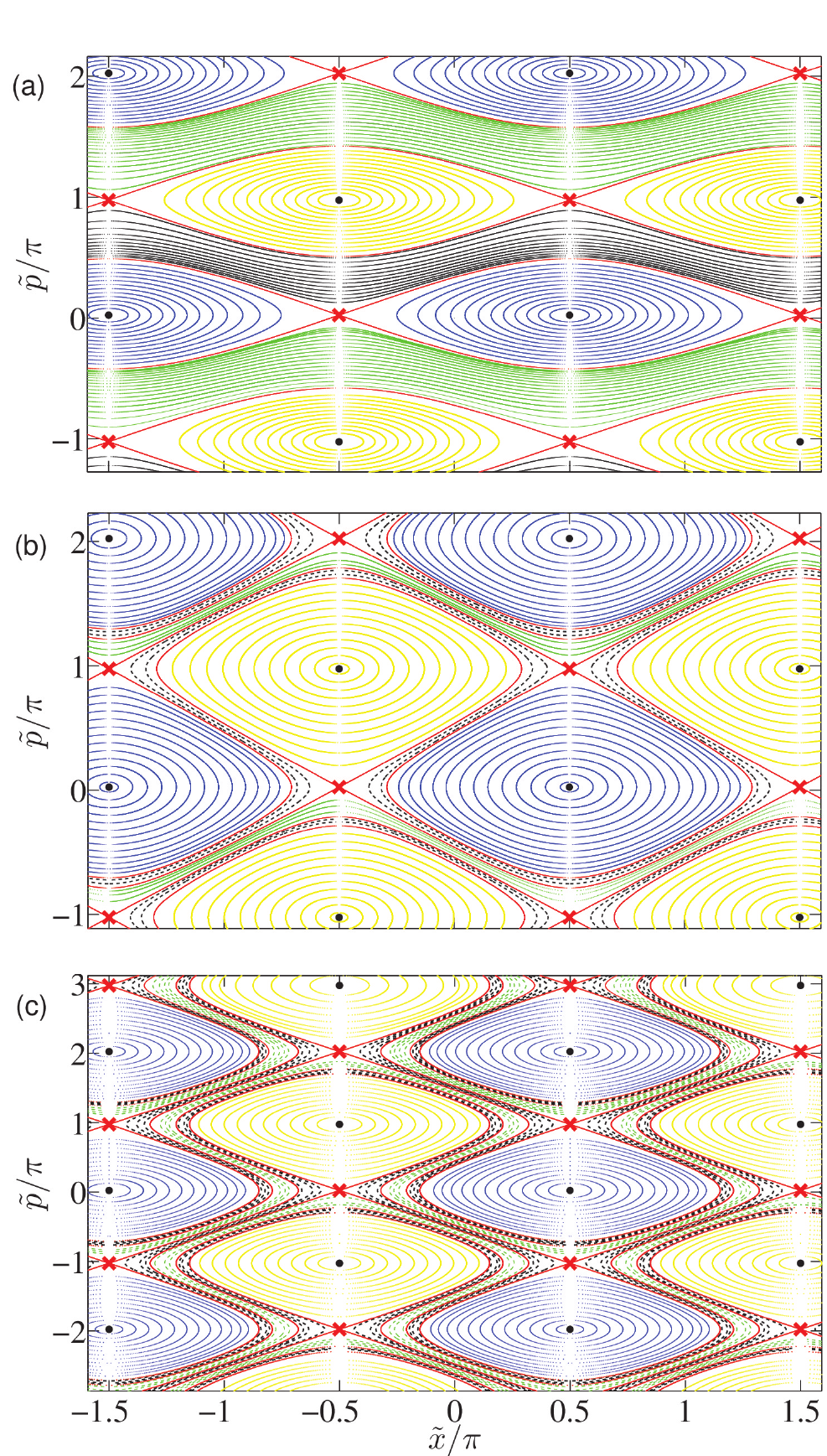}
      \caption{(color online). Phase portraits of the dynamical system (\ref{eq:all-tilde})
for (a) $U=1.5$ meV ($U<U_{cr1}$); (b)  $U=3.45$ meV ($U>U_{cr1}$) and (c) $U=4.25$ meV ($U>U_{cr2}$).
The positions of equilibrium points are indicated by black solid circles (centres) and red crosses (saddles). The localized orbits are represented by blue and yellow dots. Smooth solid  curves correspond to the unbounded trajectories, dashed curves mark the meandering trajectories, and red curves denote the separatrices.        
   } \label{fig:ph_sp}
 \end{figure}

For $v_0>v_s$, which is  valid for typical SLs, the dynamical system (\ref{eq:all-tilde})
demonstrates a countable set of equilibrium points, which satisfy the following conditions
\begin{subequations}
\label{eq:all-fp}
\begin{eqnarray}
v_0\sin\tilde{p} =v_s,\label{eq:fpx}\\
\cos\tilde{x}=0.  \label{eq:fpp}
\end{eqnarray}
\end{subequations}
Equations (\ref{eq:all-fp}) 
evidence that a steady state corresponds to the situation, when an electron moves with the velocity of the acoustic wave $v_s$ being at the position coinciding with one of extrema of the potential $V(x,t)$. This yields the following coordinates of the fixed points 
\begin{subequations}
\label{eq:all-fp_tilde}
\begin{eqnarray}
\tilde{x}&=&\frac{\pi}{2} +m\pi,  \label{eq:fp_xtilde}\\
\tilde{p}&=&(-1)^{l} \sin^{-1}\left(\frac{v_s}{v_0}\right) +l\pi, \label{eq:fp_ptilde}
\end{eqnarray}
\end{subequations}
where $m$ and $l$ are arbitrary integer numbers.
A simple stability analysis reveals that all these fixed points 
are always either centres or saddles. These points are periodically spread in the phase space as it is illustrated 
in Fig.~\ref{fig:ph_sp}, where the black circles denote the centres, and the red crosses indicate positions of the hyperbolic fixed points (saddles).

We will consider significance of the hyperbolic points  in our forthcoming analysis of global bifurcations, but before that we need to review evolution of the phase portraits with a variation of the acoustic wave amplitude $U$.

After dividing (\ref{eq:ptilde}) by (\ref{eq:xtilde}) and performing integration we obtain the phase trajectory equation 
\begin{eqnarray}\nonumber
\tilde{x}=(-1)^j \sin^{-1} 
 \left\{\sin\tilde{x}_0 - \frac{\hbar v_s}{Ud}\left[\frac{v_0}{v_s}( \cos\tilde{p}\right. \right. \\- 
 \Biggl. \Biggl. \cos\tilde{p}_0) +\tilde{p}-\tilde{p}_0\bigg]\bigg\} +j\pi, \label{eq:pp}
 \end{eqnarray}
where $j$ is an integer number, and  ($\tilde{x}_0$,$\tilde{p}_0$) is an initial condition.
Figure \ref{fig:ph_sp} displays the phase portraits of the dynamical system  (\ref{eq:all-tilde})
calculated using Eqs.~(\ref{eq:all-fp_tilde}) 
and (\ref{eq:pp}) 
for three characteristic values of $U$.
Although the physical meaning of the both variables $\tilde{x}$  and $\tilde{p}$ allows their wrapping  into the interval $(-\pi,\pi]$,  we found more 
convenient to present the phase portraits in the unwrapped phase space. For small $U<U_{cr1}\approx3.1$ meV [Fig.~\ref{fig:ph_sp}(a)] the phase space of the system is represented by periodic ``islands'' of localized trajectories (blue and yellow closed orbits), which rotate around corresponding centres (black solid circles). Depending on the particular location in the phase space the localized trajectories rotate either clockwise (blue orbits) or anticlockwise (yellow orbits). The islands of the localized trajectories are intermitted with the areas of unbounded trajectories, which propagate either to positive or negative direction along the $\tilde{x}$--axis, depending on the initial value $\tilde{p}_0$. 
In the particular case of  Fig. \ref{fig:ph_sp}(a), the unbounded trajectories above the islands of the clockwise orbits  (black lines) propagate in the positive direction of $\tilde{x}$, whereas the trajectories below these islands (green lines) run in the negative direction of  $\tilde{x}$.
The characteristic regions of different trajectories are separated by a heteroclinic structure (red curves), separatrix, formed by the manifolds of the saddle points (red crosses) with the same coordinate $\tilde{p}$. 

As $U$ increases, 
the areas of the unbounded trajectories shrink, and after $U$ exceeds the critical value $U_{cr1}$, the unbounded trajectories drifting in the positive directions of $\tilde{x}$ disappear, thus manifesting  a dramatical change in the topology of the phase portrait. This phase space realignment is associated with origin of a new type of the phase trajectories, which demonstrate a  meandering behaviour [black dashed curves in Fig. \ref{fig:ph_sp}(b)] encompassing the islands of the localized orbits. Independently on the initial conditions these meandering trajectories always drift in the negative direction of  $\tilde{x}$. Further increase of  $U$  above $U_{cr2}\approx3.92$ meV eliminates completely the unbounded trajectories, but  gives rise to complex meandering trajectories, which are more elongated in $\tilde{p}$-direction and envelop more islands of the localized orbits [black dashed curves in Fig. \ref{fig:ph_sp}(c)]. These new elongated phase trajectories coexist with less elongated trajectories [green dashed curves in Fig. \ref{fig:ph_sp}(c)] that were born when $U$ exceeded $U_{cr1}$. We found out that other 
critical values of $U$ ($U_{cr3}\approx4.75$ meV and $U_{cr4}\approx5.58$ meV) indicated in Fig. \ref{fig1}(b) relate to the appearance of new meandering trajectories, which envelop  along $\tilde{p}$ -- direction a larger number of the islands of the localized orbits.
\begin{figure}
\includegraphics[scale=0.16]{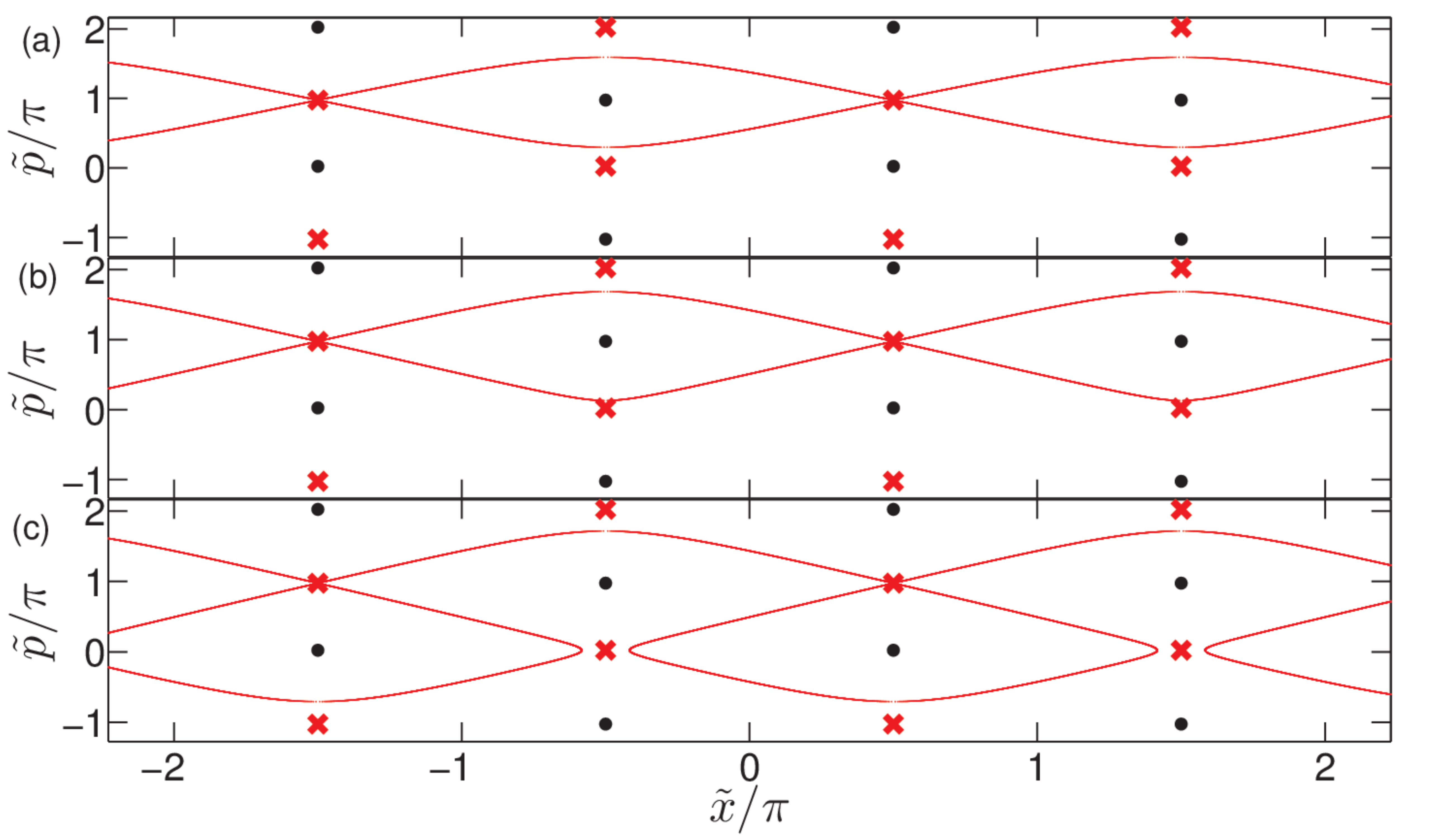}
   \caption{
(color online). Evolution of the separatrix structure with a variation of $U$ around the first bifurcation point ($U_{cr1}=3.1$ meV): 
(a) $U=2.5$ meV; (b)  $U=3$ meV; (c) $U=3.15$ meV.
The positions of the centres are indicated by black solid circles and the saddles -- by red crosses.
   \label{seprx}
   }
 \end{figure}

Such restructuring of the phase portraits is usually associated with the development of global bifurcations (instabilities) in the 
dynamical
system \cite{Kuz_bif_th}. In order to get deeper insight into onset of these bifurcations, we 
analyse an evolution of the separatrices near the critical values of $U$. 
Figure~\ref{seprx} presents the separatrix formed by  the manifolds of the saddles with coordinate 
$\tilde{p}_1=\pi-\sin^{-1}(v_s/v_0)$  
for three characteristic values of $U$ close to $U_{cr1}$. When $U<U_{cr1}$ [Fig.~\ref{seprx} (a)] the separatrix is a heteroclinic structure that  delimits the regions of the localized orbits, see also Fig.~\ref{fig:ph_sp} (a). With an increase of $U$ the manifolds of the saddles with coordinate $\tilde{p}_1$ 
approach the saddles with coordinate $\tilde{p}_2=\sin^{-1}(v_s/v_0)$ 
[Fig.~\ref{seprx} (b)], and at $U=U_{cr1}$ touch them, thus causing the global bifurcation. Further growth of $U$ enforces a manifold reconnection, which forms the separatrix consisting of both the homoclinic and hereroclinic structures  [Fig.~\ref{seprx} (c)].  In this case, the manifolds connect the neighboring saddle points having the same coordinate $\tilde{p}_1$.  
In addition, the manifolds of each saddle form a homoclinic loop, which bounds the island of localized trajectories rotating around the centres with the coordinate $\tilde{p}_2$. 
This separatrix structure provides the conditions for emergence of the meandering trajectories depicted in Fig.~\ref{fig:ph_sp} (b) by dashed curves.

 \begin{figure}
 \includegraphics[scale=0.16]{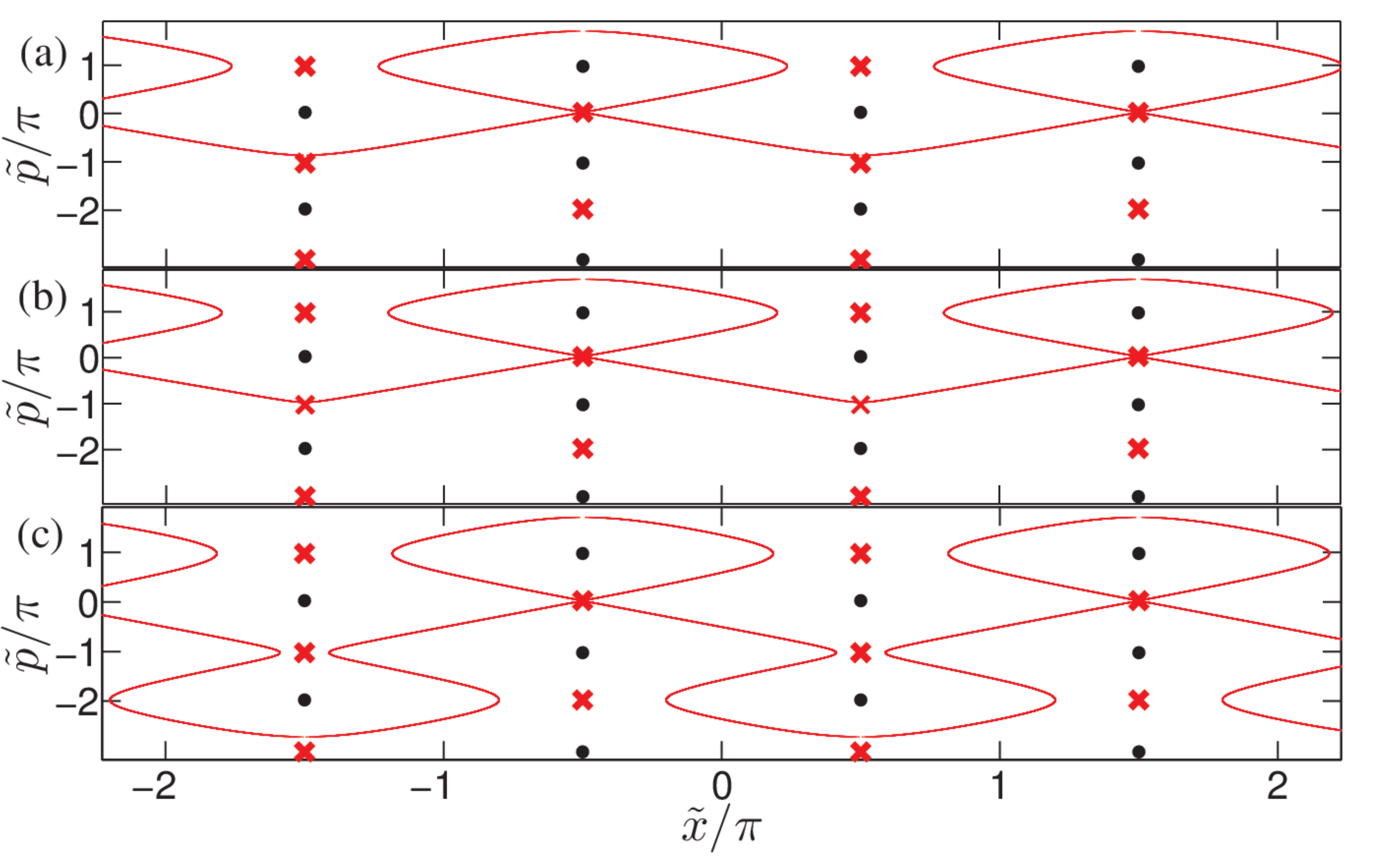}
	\caption{
(color online). Evolution of the separatrix structure with a variation of $U$ in the vicinity of the second bifurcation ($U_{cr2}= 3.92$ meV):
(a) $U=3.7$ meV; (b)  $U=3.9$ meV; (c) $U=4$ meV. Other notations are the same as in Fig.~\ref{seprx}.	
\label{seprx2}
   }
 \end{figure}  
 
Similar topological rearrangements of the phase space take place near other critical values of $U$. For example, Fig. \ref{seprx2} illustrates the evolution of the 
separatrix
topology with variation of $U$ in the vicinity of $U_{cr2}$. Before the global bifurcation [Fig.~\ref{seprx2} (a)], i.e for $U<U_{cr2}$, the separatirix is formed by a set of heteroclinic and homoclinic connections similar to one shown in Fig.~\ref{seprx} (c). However, in the present case we consider the separatrix formed by the manifolds of the saddles characterised by the coordinate 
$\tilde{p}_2$. As $U$ approaches to the critical value $U_{cr2}$ the heteroclinic parts of the separatrix come closer to the saddle points with 
$\tilde{p}_3=-\pi-\sin^{-1}(v_s/v_0)$
[Fig.~\ref{seprx2} (b)], and at the point of the bifurcation, $U=U_{cr2}$, involves these saddles into heteroclinic connections. This nonrobust structure disconnects with further increase of $U$ forming an additional lap around the centres with coordinate 
$\tilde{p}_4=-2\pi+\sin^{-1}(v_s/v_0)$,
see Fig.~\ref{seprx2}(c). This complex wriggling shape of the separatrix promotes generation of the meandering trajectories that encompass  several areas of the localized trajectories similar to those shown in  Fig.~\ref{fig:ph_sp} (c) by dashed curves. 
We also reveal that all other critical values of $U$ associated with the local extrema of the dependence $v_m(U)$ [see Fig.~\ref{fig1}(b)] correspond to additional global bifurcations, which involve new saddle points into heteroclinic connections. 

Since the global bifurcations are attributed to the situations, when a manifold of one saddle touches another saddle, 
the conservation of energy $H'=\text{const}$ [Eq.~(\ref{eq:Eprime})] can be used for finding the bifurcation points
analytically:
\begin{equation}
U_{cr}=\frac{\hbar}{d}\frac{v_0\left(\cos\tilde{p}_{s i}-\cos\tilde{p}_{s j}\right)+
v_s\left(\tilde{p}_{si}-\tilde{p}_{sj}\right)}{\sin\tilde{x}_{sj}-\sin\tilde{x}_{si}}, 
\label{eq:u_cr-step1}
\end{equation}
where $(\tilde{x}_{si},\tilde{p}_{si})$ and $(\tilde{x}_{sj},\tilde{p}_{sj})$ are coordinates of a pair of saddles involved in a given bifurcation as we described above in our explanation of Figs.~\ref{seprx} and \ref{seprx2}  
\footnote{Alternatively, the pairs $(\tilde{x}_{si},\tilde{p}_{si})$ and $(\tilde{x}_{sj},\tilde{p}_{sj})$
involved in the equation (\ref{eq:u_cr-step1}) can include any combinations of the stationary points from the whole set 
(\ref{eq:all-fp_tilde}).
In such a way elliptic points will be automatically excluded from the consideration as not connected by any phase trajectories, and only proper combinations
of two hyperbolic points can satisfy the conservation of energy for every value of $U_{cr}$. 
}.
Next, using  these coordinates we obtain the following explicit expression for critical values of $U$:
\begin{eqnarray}
U_{cr_n}&=&\frac{\hbar v_s}{d}\bigg[\sqrt{\left(\frac{v_0}{v_s}\right)^2-1} \nonumber \\
&+&\sin^{-1}\left(\frac{v_s}{v_0}\right)+\left(n-\frac{3}{2}\right)\pi\bigg]\quad (n\geq 1).
\label{eq:u_cr}
\end{eqnarray}
For our choice of SL parameters this equation gives  
$U_{cr1}$=3.1 meV, $U_{cr2}=3.92$ meV, $U_{cr3}=4.75$ meV, and $U_{cr4}$=5.58 meV, which are in an excellent agreement with critical values found in numerical simulations and shown in Fig. \ref{fig1} (b).

The criterion (\ref{eq:u_cr}) constitutes the main analytical result of our work.  
Physical processes behind this formula are generally related to resonances in a transition scattering of the sound wave by electrons \cite{Ginzburg_bristol} placed in the SL periodic potential. 
More detailed analysis of theses processes in the relation to the global bifurcations will be published elsewhere.

\subsection{\label{subsec:level-AA} Typical dynamical regimes and real space trajectories }

\begin{figure}
   \includegraphics[scale=0.9]{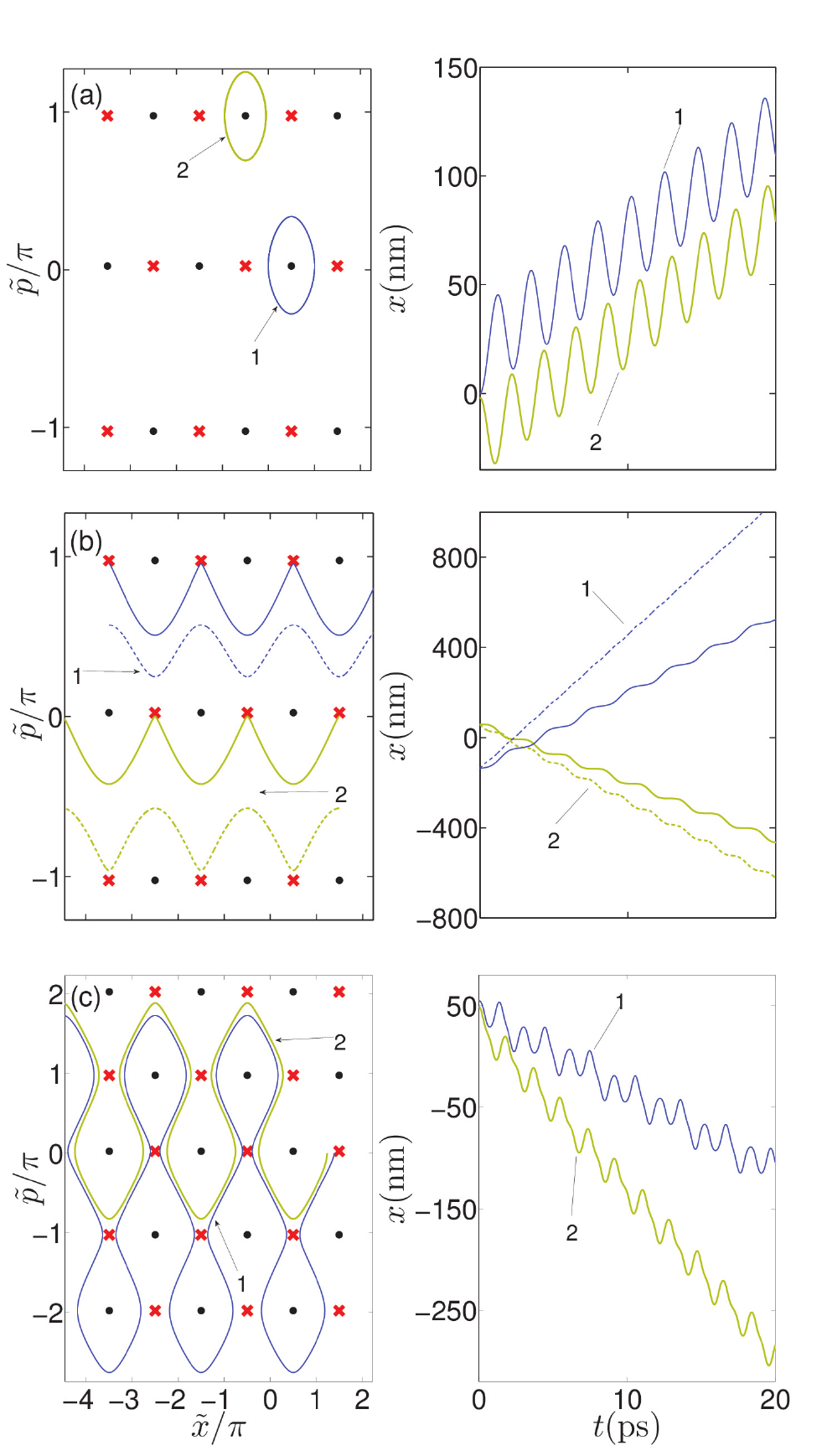}
   \caption{(color online). Three distinct types of the phase trajectories (left panels) and the corresponding electron trajectories in the real space (right panels) for (a) $U=1.5$ meV; (b) $U=2.5$ meV; (c) $U=4.2$ meV.  
	\label{real_tr}
   }
 \end{figure}

Analysis of the phase portrait changes with variation of $U$ allows us to reveal three characteristic types of the phase trajectories that determine the electron transport in the real space. 
Typical examples of those trajectories and the related electron trajectories in the real space are summarized in Fig. \ref{real_tr}.
Figure \ref{real_tr} (a) displays the localized phase trajectories, which orbit around the centres either clockwise (trajectory 1 in the left panel) or anticlockwise (trajectory 2 in the left panel) in dependence on the initial conditions. However, independently of the rotation directions these phase trajectories manifest
themselves in the drift of the electrons with speed $v_s$ towards the positive direction of $x$. This drift is accompanied by periodic oscillations, see the right panel of Fig. \ref{real_tr}(a). In this regime the acoustic wave just drags electrons that are localized within a single minimum of the acoustic potential \cite{greenaway2010using}. 

For small $U$ the localized phase trajectories coexist with the unbounded trajectories, which are illustrated in the left panel of Fig. \ref{real_tr} (b). The direction of the propagation of the unbounded trajectories along $\tilde{x}$-axis  depends on the initial conditions. For instance, in the left panel of Fig. \ref{real_tr}(b) the blue trajectories (curves 1) run towards the positive directions of $\tilde{x}$, whilst the yellow trajectories (curves 2) move in the negative direction along $\tilde{x}$. In this regime $\tilde{p}$  only slightly oscillates around a certain mean value. Therefore, according to (\ref{xdot}) the time-averaged velocity of the electron in this regime can be estimated as $\alpha v_0$, where $\alpha$ is  a constant constituting the averaged value of $\sin \tilde{p}$, either positive or negative. The electrons trajectories for this type of dynamics are shown in the right panel of Fig. \ref{real_tr} (b).  One can see that the electrons are not trapped by the acoustic wave, and can move either in positive or negative directions, depending on the initial value of $\tilde{p}$.  Since the initial conditions directly affect constant $\alpha$, they influence the time-averaged electron velocity. This is clearly seen in Fig. \ref{real_tr} (b), where few trajectories are depicted for comparison. 

As $U$ grows the amplitude of nonlinear $\tilde{p}$ - oscillations becomes larger and eventually exceeds the size of the first Brillouin zone 
$|\tilde{p}|>\pi$ resulting in Bragg reflections of the electron
and the rise of complex Bloch oscillations.
In terms of nonlinear dynamics this indicates a global bifurcation, which gives birth to the meandering trajectories 
of the type shown in the left panel of Fig. \ref{real_tr} (c). 
For $U>U_{cr2}$, depending on the initial conditions, the meandering phase trajectory can encompass either even (blue curve 1) or odd (yellow curve 2) number of the islands of the localized trajectories [see also black  or green dashed curves in Fig.~\ref{fig:ph_sp}(c)]. 
Nevertheless, in both cases these meandering trajectories wander towards the negative directions of $\tilde{x}$. Such phase space dynamics determines the electron trajectories rapidly moving in the negative direction as shown in the right panel of Fig.~\ref{real_tr} (c).

\subsection{\label{subsec:level-B} Directed transport in terms of phase trajectories}
 
The contribution of different phase trajectories to electron transport in SL explains well  the shape of the dependences $v_m(U)$ and $v_d(U)$ in Fig. \ref{fig1}.
As it was pointed out in  Section \ref{sec:level1}, $v_m$ can be understood as a  time-averaged electron velocity additionally averaged over initial positions $x_0$.  
Since we consider the case close to zero temperature, a zero initial momentum $p_0$ was assumed for all trajectories. Figure \ref{fig:ph_sp}(a) evidences that for a weak acoustic wave only localized [Fig. \ref{real_tr}(a)] and unbounded trajectories [Fig. \ref{real_tr}(b)]  determine the value of $v_m$.  With this, for zero initial momenta the unbounded trajectories generate backward motion of the electrons, which competes with positive drift promoted by the localized trajectories. 
Since for the for the given SL parameters the ratio $v_s/v_o$ is small enough, the positions of  saddles, according to (\ref{eq:fp_ptilde}), are very close to $\tilde{p}=0$.
The measure of localised  trajectories started with zero momenta is much larger than the one for unbounded trajectories, which predetermines the positive value of $v_m$. Moreover, as $U$ increases, the area of the islands of the localized trajectories increases as well, diminishing the proportion of unbounded trajectories. This explains a rapid growth of $v_m$ followed by its saturation within the range of $U$ between  0 and $U_{cr1}$. For $U>U_{cr1}$ the meandering trajectories  [black dashed curves in Fig. \ref{fig:ph_sp}(b)], which  promotes rapid backward motion of electrons [Fig. \ref{real_tr}(c)], start to affect $v_m$. Increase of $U$ widens the area of the meandering trajectories on the phase plane, thus evoking a sharp drop of $v_m$. The next global bifurcation at $U=U_{cr2}$ creates a homoclinic loop [Fig. \ref{seprx}(c)], which bounds the localized trajectories. Increase of $U$ expands this loop, thus widening area of the localized trajectories and shrinking the region of meandering trajectories. These changes in the phase portraits slightly increase $v_m$.  Further global bifurcations  repeat the scenario described above, and produce additional extrema in the graph of $v_m(U)$ shown in Fig. \ref{fig1} (b).  Essentially same arguments are  also valid for explanation of the dependence $v_d(U)$ presented in Fig. \ref{fig1} (a). However, scattering events limit the length of the trajectories contributing to the drift velocity. In Eqs. (\ref{Boltz_vd}) and (\ref{vd_vn}) the reduction of the miniband velocity due to scattering  is reflected in the presence of the exponent involving the scattering time $\tau$. These factors weaken the effect of the phase portraits restructuring on the drift velocity, and therefore the manifestation of only two most prominent global bifurcations is visible in the profile of $v_d(U)$.

\begin{figure}[t]
 \includegraphics[scale=0.13]{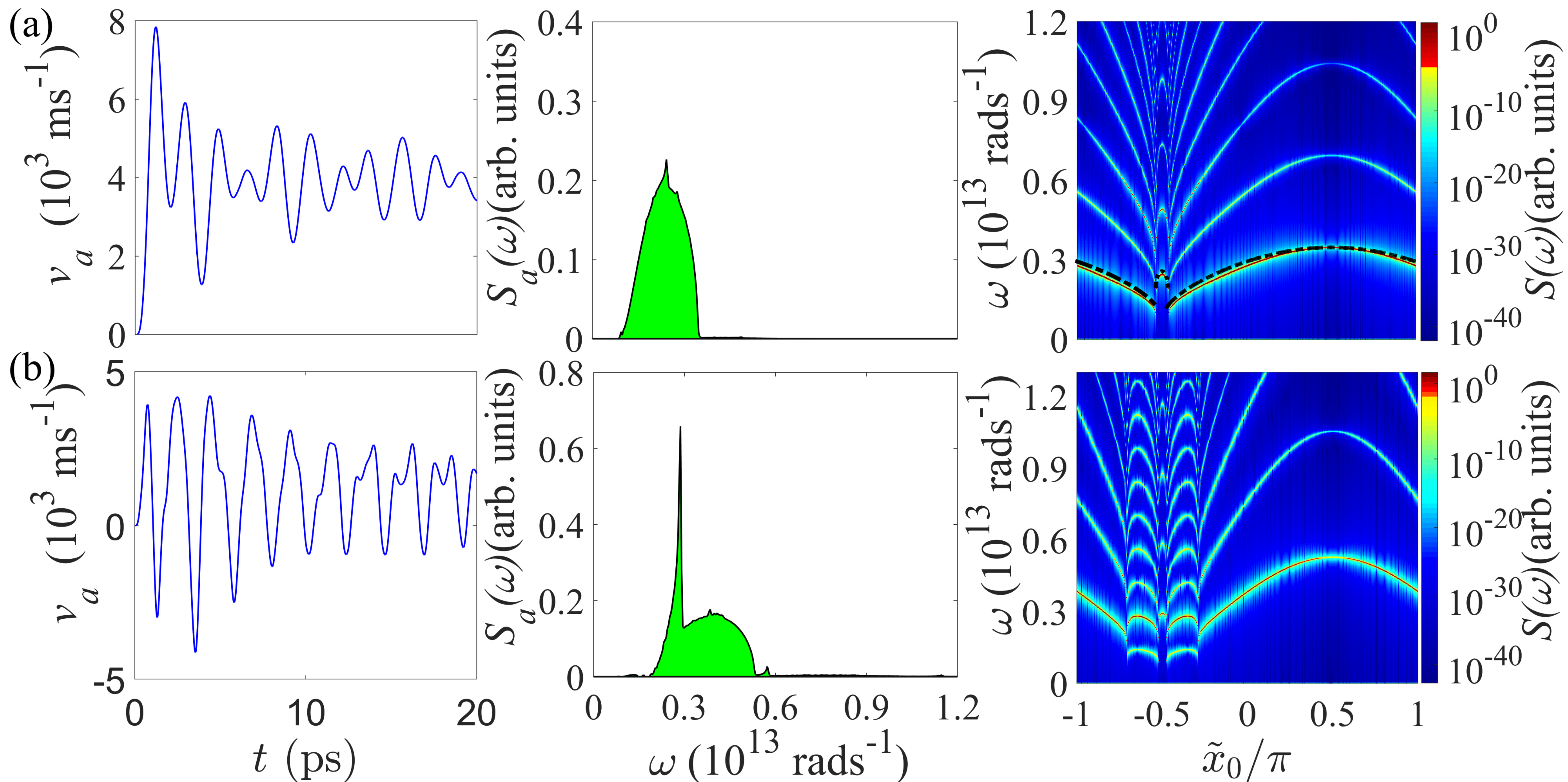}
 \caption{(color online). A typical time-realization of $v_a(t)$ (left panel), its Fourier spectrum $S_a(\omega)$ (middle panel) and the dependence of the spectrum $S(\omega)$ on the initial position of the particle $x_0$ (right panel) calculated for (a) $U=1.5$ meV $<U_{cr1}$   and (b) $U=3.45$ meV $>U_{cr1}$. Dot-dashed lines mark the dependence $\omega_1(x_0)$ calculated analytically (see the text for details). The spectra in the right panels are given in log-scale in order to make the spectral peaks more noticeable.} 
 \label{spec1}
 \end{figure}

\section{\label{sec:level4}SPECTRAL ANALYSIS OF TRAJECTORIES}

Previously, we showed that  electrons starting from different initial conditions can move along topologically different trajectories, see Figs. \ref{fig:ph_sp} and \ref{real_tr}. Therefore, in order to study the frequency characteristics of the global transport, we introduce $x_0$-averaged velocity of electrons $v_a(t)=\langle\dot{x}\rangle_{x_0}$ and its Fourier spectrum $S_a(\omega)$. To calculate $\langle\dot{x}\rangle_{x_0}$ we average $\dot{x}(t)$ for an ensemble of electron trajectories 
with different $x_0$ from the interval [$-\lambda/2, \lambda/2$). We assume all electrons start with the same $p_0=0$ (low temperature limit). Typical time realisations of $v_a(t)$  in the vicinity of $U=U_{cr1}$ are displayed in the left panels of Fig. \ref{spec1}. The left panel of Fig. \ref{spec1} (a) presents a realisation of $v_a(t)$ calculated for  
$U=1.5$ meV $<U_{cr1}$. It demonstrates oscillations, whose amplitude changes in prominently erratic manner. 
These oscillations are characterised by a broadband spectrum $S_a(\omega)$ [middle panel of Fig. \ref{spec1}(a)] centred around frequency $\omega=2.5\times10^{12}$ rad/s$^{-1}$. 
Their broadband character is determined by contribution of the localized  and unbounded
trajectories shown in Fig. \ref{fig:ph_sp}(a). 

The above  localized  and unbounded trajectories can be quite accurately described within a pendulum approximation. Indeed, by assuming a small change of $\tilde{p}$, Eqs.~(\ref{eq:all-tilde}) can be reduced to the pendulum equation
\begin{equation}
\frac{d^2\tilde{y}}{d\tilde{t}^2}+\Omega^2\sin \tilde{y} =0. 
\label{eq:pend}
\end{equation}
Here $\tilde{y}=\tilde{x}-\pi/2$, and $\Omega=(U\Delta/2)^{1/2}(d/\hbar v_s)$. The spectrum of oscillations of the system (\ref{eq:pend}) is known to depend on the integral of motion 
${\cal H}=\dot{\tilde{y}}^2_0/2-\Omega^2\cos\tilde{y}_0$,
where $\tilde{y}_0$, and $\dot{\tilde{y}}_0$ are the initial values of $\tilde{y}$ and $d\tilde{y}/d\tilde{t}$, respectively. 
In particular, the position of the the first harmonic $\omega_1$ in the spectra of the oscillations $\tilde{y}(t)$
can be expressed as $\omega_1=\pi\Omega\omega_s/[2K(\kappa)]$ for localized trajectories, and as $\omega_1=\pi\Omega\omega_s \kappa/K(1/\kappa)$ for unbounded trajectories 
\cite{Chirikov79}. 
Here $\kappa^2=1/2+{\cal H}/(2\Omega^2)$, and $K(.)$ is the complete elliptic integral of the first kind.  

To illustrate this better, we calculate the spectra of individual trajectories, $S(\omega)$, and compare them with $S_a(\omega)$.
The right panel in Fig. \ref{spec1} (a) displays the dependence of the spectrum, $S(\omega)$, of $\dot{x}(t)$ on initial condition $\tilde{x}_0$ calculated numerically using basic Eqs.~(\ref{eq:all-dot}). 
The colormap denotes the different values of $S(\omega)$,
while the dot-dashed lines show the frequency $\omega_1$ found analytically. The figure reveals that the frequency of the most prominent spectral peak changes  significantly  with variation of $x_0$, having a mean value close to the centre of the spectral band of $S_a$ [cf. middle panel of Fig. \ref{spec1} (a)]. This is also  confirmed  by dependence $\omega_1(x_0)$ calculated analytically [dot-dashed curve in the right panel of Fig. \ref{spec1}(a)], which demonstrates an excellent agreement with the spectra calculated numerically.

When $U$ slightly exceeds $U_{cr1}$, the averaged oscillations $v_a(t)$ become less erratic [left panel of  Fig. \ref{spec1} (b)], which manifests also in the appearance of a pronounced peak in $S_a(\omega)$  [middle panel of  Fig. \ref{spec1} (b)]. These changes in the spectrum are associated with the emergence of the meandering  phase trajectories [Fig. \ref{fig:ph_sp}(b)] corresponding to the frequency-modulated miniband velocity $\dot x$(t).  This type of trajectories represents the complex Bloch oscillations, and can be viewed as a  frequency modulated signal with the characteristic cut-off frequency  $\omega_{cut}=k_sUd/\hbar$ \cite{greenaway2010using}. Notably, for $v_0\gg v_s$, 
$U_{cr_1}\approx\Delta/2$ and therefore when $U\approx U_{cr1}$
\begin{equation}
\frac{\omega_{cut}}{\omega_s}\approx\frac{v_0}{v_s}\gg 1.
\label{eq:cutoff}
\end{equation}
At the same time,  the localised trajectories condition  the broadness of the spectrum $S_a(\omega)$ presented in the middle panel of  Fig. \ref{spec1} (b). 

Remarkably, the frequency of the dominant peak of the frequency-modulated oscillations is weakly depended on the initial conditions.  The dependence of the spectrum $S(\omega)$ of $\dot{x}(t)$ on the initial position $x_0$ is shown in the right panel of Fig. \ref{spec1} (b). The figure confirms that for the meandering trajectories starting from  the vicinity of  $\tilde{x}_0=-\pi/2$ [see also Fig. \ref{fig:ph_sp}(b)], the position of the  dominant peak in $S(\omega)$ changes weakly. 
Moreover, the positions of the most prominent peaks in the spectra $S(\omega)$ [right panel of Fig. \ref{spec1} (b)], corresponding to different meandering trajectories, are well agreed with the position of the dominant peak in $S_a(\omega)$ (middle panel). All this leads to regularization of $v_a(t)$ oscillations [see right panel of Fig. \ref{spec1} (b)].

Thus, for $U>U_{cr1}$ a propagating sound wave induces a high-frequency response  in ballistic transport of electrons, which is characterised by a sharp pronounced spectral peak, whose frequency  for the given parameters
about ten times exceeds the frequency of the acoustic wave. The emergence of this peak is associated with generation 
of complex Bloch oscillations, which are represented in the phase plane  by the meandering trajectories. We also analysed the higher-order bifurcation at $U_{cr_i}$, and show that 
the variation of the wave amplitude $U$ allows to considerably tune the frequency of the main peak, and to control its height (see Appendix \ref{App1} for more details).
 
\section{\label{sec:level5}CONCLUSION}

Combining nonperturbative methods of nonlinear dynamics and kinetics,
we disclosed the bifurcation mechanisms governing the miniband electron transport  induced in a periodic superlattice potential by a propagating acoustic wave. Analysis of the phase portraits in the moving reference frame allowed us to identify the specific bifurcations, which are developed with variation of the wave amplitude $U$. These global bifurcations, evoking a structural rebuilding of the phase space, cause a sudden change of both the drift velocity $v_d$ and the time-averaged velocity $v_m$ of electrons. We analytically estimated the critical values of $U$, corresponding to the bifurcations, by deriving the conditions for appearance of the heteroclinic connections in the phase space of the system. Both dependences $v_d(U)$ and $v_m(U)$ demonstrate prominent maxima followed by an abrupt drop. The values of $U$ corresponding to these specific features in the velocity dependences are in excellent agreement with the critical $U$ calculated analytically.

We also revealed and classified three characteristic types of ballistic electron trajectories generated by an acoustic wave.  They are attributed to (i)  motion of electrons confined by a propagating potential wave (localized trajectories in the moving reference frame), (ii) unconfined electron motion (unbounded trajectories in the moving reference frame) and (iii) complex Bloch oscillations (meandering trajectories in in the moving reference frame). Depending on the particular value of $U$ the contribution from each type of trajectories to the charge transport is different. This predefines both drift and frequency properties of the transport regime realized in the system.  By choosing an appropriate amplitude of the acoustic wave one can generate oscillations of the averaged electron velocity 
$v_a(t)$, which will be characterised either by a broadband spectrum or by a spectrum with a pronounced peak. With this, the central frequency of the velocity spectrum can significantly exceed the frequency of the propagating wave and can be controlled by a variation of $U$.

Our results suggest that fast miniband electrons driven by an acoustic wave should spontaneously emit submillimeter electromagnetic waves. 
The waves could be detected by the technique of time-resolved THz-emission spectroscopy, in the way similar to used in other radiating superlattice devices \cite{Dekorsy95,Bauer02}. 
Moreover,  the spectral signatures of the electron velocity unravel the potential of the
acoustically pumped superlattices for an amplification of high-frequency electromagnetic signals involving mechanisms similar to the 
stable Bloch gain \cite{Ktitorov72,hyart2009terahertz}.
From nonlinear dynamics perspective, our semiclassical tight-binding model belongs to the so-called driven Harper models 
\cite{Kolovsky12}. 
In this respect, it interesting to compare possible physical manifestations of the global bifurcations in different nonlinear systems of this class \cite{Kolovsky12,Iomin98,Petschel91}.
At last but not least, the model of a superlattice driven by a plane wave is potentially realizable with ultracold atomic matter waves \cite{greenaway2013resonant}. It provides an opportunity to model the discussed nonlinear electro-acoustic effects  utilizing cold atom manipulations, thus contributing to rapidly developing areas of atomtronics \cite{seaman2007atomtronics,Ryu15} and the related simulations of lattice transport phenomena \cite{Meier16}.

\section{Acknowledgments}
This work has been supported by EPSRC (grant EP/M016099/1). AGB acknowledges support from the the Russian Science Foundation (grant 14--12--00222).
\appendix

\section {\label{App1}Spectral analysis of higher-order bifurcations}

\begin{figure}[t]
  \includegraphics[scale=0.75]{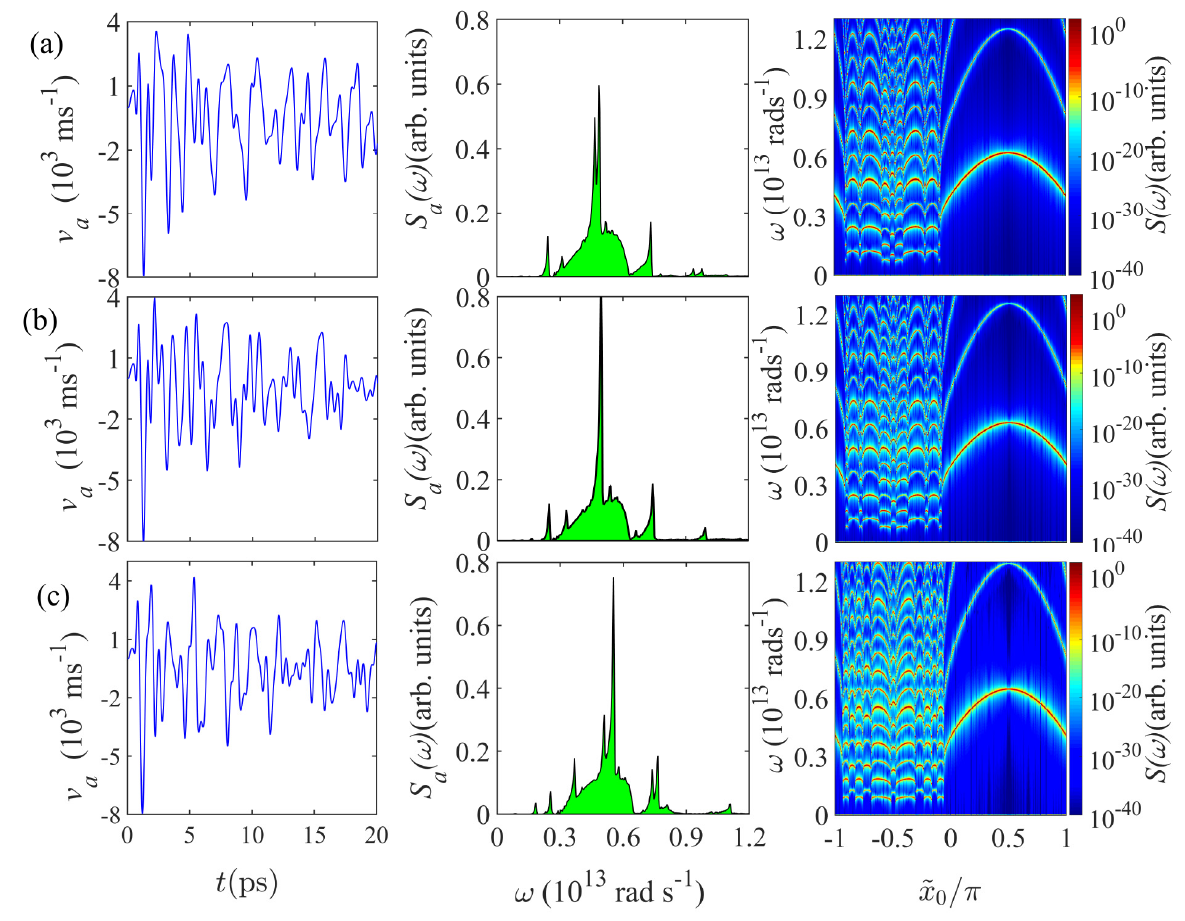}
   \caption{(color online). A typical time-realization of $v_a(t)$ (left panel), its Fourier spectrum $S_a(\omega)$ (middle panel) and the dependence of the spectrum $S(\omega)$ on the initial position of the particle $x_0$ (right panel) calculated for (a) $U=4.85$ meV, (b) $U=4.932$ meV, and (c)  $U=5.2$ meV. The spectra in the right panels are given in log-scale in order to make the spectral peaks more noticeable. 
}
\label{spec2}  
 \end{figure}

Figure \ref{spec2} illustrates how the average velocity $v_a(t)$ and the related spectra change with variation of $U$ between two successive global bifurcations. In particular, we consider three values of $U$ between $U_{cr3}$ and $U_{cr4}$, namely, $U=4.85$ meV (a), $U=4.932$ meV (b) and $U = 5.2$ meV (c). Since for these control parameter values the unbounded trajectories do not exist, the transport is determined only by the contribution of the localised and the meandering trajectories. The latter have a complex shape encompassing three or four islands of stability. In addition, the interval of the initial conditions $x_0$ for meandering trajectories becomes comparable with the range of initial conditions for localized trajectories.
All this reflects in the dynamics of the average velocity of particles $v_a(t)$, which now  demonstrates very complicated behaviour (left panels). As a result, the spectrum of $v_a$ becomes broader and reacher due to generation of new spectral components (compare middle panels of Figs. \ref{spec1} and \ref{spec2}). However, for all values $U$ considered a pronounced peak persists in the spectra, having a frequency significantly larger than for the case of $U_{cr1}<U<U_{cr2}$ [Fig. \ref{spec1}(b)].  The specific content of $S_a(\omega)$ can be understood after inspection of the dependence of the velocity spectrum $S(\omega)$ on the initial position $\tilde{x}_0$ shown in the right panel of Fig. \ref{spec2}. Here, a significant range of the initial positions corresponds to the meandering trajectories. Since each meandering trajectory implies a frequency-modulated electron oscillations, they contribute to $S_a(\omega)$ by multiple peaks typical for frequency modulated signals. With this, the spectral components of  $S(\omega)$ only slightly depend on the initial position $\tilde{x}_0$. This promotes quite prominent peaks in the spectrum of the averaged velocity $S_a(\omega)$. The localized trajectories in their turn provide a broad-band background situated in the range 3$\times10^{12}$ -- 6$\times10^{12}$ rad/s. As $U$ changes between $U_{cr3}$ and $U_{cr4}$, the frequency of the most prominent peak of $S_a(\omega)$ slightly increases. Its height changes non-monotonically attaining the maximum at the value of $U\approx4.93$ meV corresponding to the local maximum of $v_m$ in Fig. \ref{fig1}(b).

\bibliography{biblio11K.bib}
\end{document}